%% file: main.tex
\documentclass[screen,authorversion, nonacm]{acmart}
\usepackage{enumitem}
\usepackage{subcaption}

\AtBeginDocument{%
  }

\setcopyright{acmlicensed}
\copyrightyear{2018}
\acmYear{2018}
\acmDOI{XXXXXXX.XXXXXXX}
\acmConference[Conference acronym 'XX]{Make sure to enter the correct
  conference title from your rights confirmation email}{June 03--05,
  2018}{Woodstock, NY}
\acmISBN{978-1-4503-XXXX-X/2018/06}

\begin{document}

\title[Design Considerations for Human Oversight]{Design Considerations for Human Oversight of AI: Insights from Co-Design Workshops and Work Design Theory}

\author{Cedric Faas}
\orcid{0009-0000-7918-4233}
\email{faas@cs.uni-saarland.de}
\orcid{1234-5678-9012}
\affiliation{
  \institution{Saarland Informatics Campus, Saarland University}
  \city{Saarbrücken}
  \country{Germany}
}

\author{Sophie Kerstan}
\orcid{0000-0002-0805-6485}
\email{sophie.kerstan@psychologie.uni-freiburg.de}
\affiliation{
  \institution{Department of Psychology, University of Freiburg}
  \city{Freiburg im Breisgau}
  \country{Germany}
}

\author{Richard Uth}
\orcid{0000-0001-9697-2311}
\email{richard.bergs@psychologie.uni-freiburg.de}
\affiliation{
  \institution{Department of Psychology, University of Freiburg}
  \city{Freiburg im Breisgau}
  \country{Germany}
}

\author{Markus Langer}
\orcid{0000-0002-8165-1803}
\email{markus.langer@psychologie.uni-freiburg.de}
\affiliation{
  \institution{Department of Psychology, University of Freiburg}
  \city{Freiburg im Breisgau}
  \country{Germany}
}
\author{Anna Maria Feit}
\email{feit@cs.uni-saarland.de}
\orcid{0000-0003-4168-6099}
\affiliation{
  \institution{Saarland Informatics Campus, Saarland University}
  \city{Saarbrücken}
  \country{Germany}
}

\renewcommand{\shortauthors}{Faas et al.}

\begin{abstract}
As AI systems become increasingly capable and autonomous, domain experts’ roles are shifting from performing tasks themselves to overseeing AI-generated outputs. Such oversight is critical, as undetected errors can have serious consequences or undermine the benefits of AI. Effective oversight, however, depends not only on detecting and correcting AI errors but also on the motivation and engagement of the oversight personnel and the meaningfulness they see in their work. Yet little is known about how domain experts approach and experience the oversight task and what should be considered to design effective and motivational interfaces that support human oversight. To address these questions, we conducted four co-design workshops with domain experts from psychology and computer science. We asked them to first oversee an AI-based grading system, and then discuss their experiences and needs during oversight. Finally, they collaboratively prototyped interfaces that could support them in their oversight task.
Our thematic analysis revealed four key user requirements: understanding tasks and responsibilities, gaining insight into the AI’s decision-making, contributing meaningfully to the process, and collaborating with peers and the AI. We integrated these empirical insights with the SMART model of work design to develop a generalizable framework of twelve design considerations. Our framework links interface characteristics and user requirements to the psychological processes underlying effective and satisfying work. Being grounded in work design theory, we expect these considerations to be applicable across domains and discuss how they extend existing guidelines for human–AI interaction and theoretical frameworks for effective human oversight by providing concrete guidance on the design of engaging and meaningful interfaces that support human oversight of AI systems.
\end{abstract}


\keywords{Do, Not, Us, This, Code, Put, the, Correct, Terms, for,
  Your, Paper}


\maketitle
\input{content/introduction}
\input{content/related_work}
\input{content/method}
\input{content/results}
\input{content/discussion}

\input{content/conclusion}
\begin{acks}
This work was partially funded by the DFG grant 389792660 as part of TRR 248 CPEC – Center for Perspicuous Computing.
\end{acks}

\section{GenAI Usage Disclosure}
The authors of this paper used generative AI for editing the manuscript, improving the quality of existing text, and creating material used in the co-design workshop. In \autoref{sec: oversight task}, we describe how the materials were generated and provide information about the prompts we used in \autoref{app:prompts}.

\bibliographystyle{ACM-Reference-Format}
\bibliography{bibliography}

\appendix
\include{content/appendix}

\end{document}

%% file: content/introduction.tex
\section{Introduction}
AI systems are increasingly capable of autonomous action and decision-making and are more and more deployed to enhance performance, improve quality, or enable new types of applications~\cite[see][for an overview]{lai_towards_2023}. However, in high-risk contexts such as medicine or in the educational or judicial system, there are practical, ethical, and legal reasons for adding one or multiple layers of human oversight to automated AI operations. Practically, humans can serve as a fallback when AI systems malfunction or fail and can detect inaccurate or inadequate (e.g., discriminatory) outputs to prevent risks to safety, health, or fundamental human rights. Ethically, human oversight ensures accountability and preserves human judgment in decisions affecting people’s lives~\cite{jobin_global_2019, crootof_humans_2023}. Finally, legal frameworks, such as the European AI Act, mandate human oversight for high-risk applications that affect safety, health, or fundamental rights.

If human oversight is required, the conditions for it must be designed to make oversight \emph{effective}. Effective oversight means that human overseers are able to detect inaccurate or inadequate system outputs and malfunctions and intervene appropriately to prevent possible negative effects~\cite{langer_complexities_2025}. This effectiveness is determined by the sociotechnical conditions of human oversight~\cite{sterz_quest_2024}, which crucially includes the user interface that human oversight personnel use to monitor and intervene in AI operations. Prior work in explainability~\cite{ali_explainable_2023, Langer_2021, longo_explainable_2024}, and human-AI interaction design~\cite{gomez_human-ai_2025, zhao_human-ai_2025}, but also research on traditional automated systems and insights from the human factors community \cite{endsley2003designing, parker_grote_2020, cummings2006automation, sheridan2019individual, langer_effective_2024}, offer valuable insights into how such interfaces need to be designed to enable understanding, detection, and control for humans to be effective in their oversight tasks. 

However, oversight is not only a question of technical effectiveness but also of what organizational psychology calls \emph{work design}, that is ``the content and organization of one’s work tasks, activities, relationships, and responsibilities''~\cite[p. 662]{parker2014beyond} and how they affect workers’ psychological needs, well-being, and job satisfaction. 
Oversight personnel must monitor AI systems for extended periods, which can make their roles monotonous, stressful, or detached from the core of their expertise, in particular, as human oversight of AI systems might substantially shift the responsibilities of domain experts, requiring them to transition from performing tasks based on their own judgment to overseeing AI systems and intervening effectively.
Interfaces must therefore not only support accurate monitoring and timely intervention but also foster engagement, meaning, and a sense of agency~\cite{sterz_quest_2024}. Designing for human oversight therefore requires designing for ``good work'', those responsible for oversight stay engaged and motivated in their roles..

In this paper, we thus adopt a human-centered perspective on human oversight and ask: \emph{Which aspects should we consider when designing oversight interfaces to help domain experts be effective, motivated and engaged in overseeing AI systems?}
To address this question, we combined a participatory design approach with theoretical integration from work design. 
We first conducted four co-design workshops in which domain experts from different backgrounds (i.e., computer science, psychology) oversaw an AI-based grading system for student tests. Reflecting on their own approaches and experiences during the oversight task, participants discussed their needs and requirements for an oversight interface. They then critically engaged with these requirements by designing concrete paper prototypes. 

Our qualitative thematic analysis revealed that participants often approached the oversight task by checking every individual test to assess AI, an approach they found ineffective and frustrating. Approaches that participants perceived as more positive included detecting unjustly awarded points by understanding the AI's strengths and weaknesses, ensuring fair grading of borderline cases, and focusing on tests that were easy and enjoyable to grade. Reflection on their approaches revealed four key user requirements: (1) having a clear understanding of their role, responsibilities, and tasks, (2) obtaining insights on the AI's capabilities and decision process, (3) contributing meaningfully to the task process, and (4) exchanging with other oversight personnel and the AI. During the prototyping session, participants designed interface elements addressing these requirements and discussed their trade-offs and personal priorities. 

The themes and requirements we identified in our workshops showed clear parallels to the psychological work design literature, which links task and work environment characteristics to key psychological processes that shape workers’ motivation and job satisfaction. Building on these parallels, we integrated our empirical findings with the SMART model of work design~\cite{parker2024smart}, which frames Stimulation, Mastery, Agency, Relatedness, and Tolerable demands as core dimensions of good work. From this synthesis, we developed a design framework comprising twelve considerations for designing oversight interfaces that are not only functional but also support overseers’ psychological needs, job satisfaction, and long-term engagement~\cite{parker2024smart}, thus supporting ``good work''.

In summary, the main contribution of this work is a design consideration framework for AI oversight interfaces that demonstrates how interface design can help domain experts fulfill their oversight role, meet their psychological needs, and derive meaning and satisfaction from their oversight task. This framework is grounded in empirical insights from domain experts overseeing an automated grading system, the user requirements we identified, and concrete examples of interface solutions.
By integrating these findings with the SMART model of work design, the framework is broadly applicable to domains beyond grading, such as medical diagnosis and treatment planning, juristical decision-making, asylum evaluations, and many others, where human oversight of increasingly autonomous AI will become essential for responsible AI implementation. Designing effective oversight interfaces is therefore critical not only to prevent errors and support performance but also to ensure that humans remain engaged, motivated, and satisfied in their evolving roles alongside AI.

%% file: content/related_work.tex
\section{Background \& Related Work}


\subsection{Human Oversight of AI Systems}
AI systems are increasingly used to automate or support decision-making in a variety of domains, such as education \cite{rastogi_deciding_2022}, finance \cite[e.g.,][]{cau_supporting_2023, appelganc_how_2022, alufaisan_does_2021, zhang_effect_2020}, healthcare \cite[e.g.,][]{appelganc_how_2022, cabitza_painting_2023, calisto_breastscreening-ai_2022, jacobs_how_2021}, or law \cite[e.g.,][]{grgic-hlaca_human_2019, lima_human_2021, liu_understanding_2021, kahr_it_2023, wang_are_2021} (see ~\citet{lai_towards_2023} for an overview).
Human oversight of such AI systems aims to mitigate risks~\cite{mcbride_understanding_2011, enqvist_human_2023}. Broadly speaking, oversight can be considered and implemented at different stages of a systems' lifecycle (e.g. at design time, at run-time, or inspection time~\cite{sterz_quest_2024}) and at different organizational levels (e.g. human oversight or institutional oversight~\cite{green_flaws_2022, laux_institutionalised_2023}). In this paper, we focus on human oversight of AI at system run-time, as it is required more and more by policies and legislations~\cite{green_flaws_2022}
This includes both, monitoring of the running system to detect failures, unsafe behavior, or biased outcomes~\cite{sterz_quest_2024, langer_effective_2024, enqvist_human_2023}, as well as intervening in the automated process, such as overwriting decisions or changing inputs, but could also involve delegating the intervention to other stakeholders~\cite{sterz_quest_2024}. Since monitoring and intervention require expert judgment, the role of a human oversight person is typically taken by domain experts, who have the necessary expertise and background to judge the accuracy of AI outputs, or by oversight experts, who are trained to detect AI failures~\cite{sterz_quest_2024}.

\citet{sterz_quest_2024} argue that an oversight person is effective if and only if four requirements are met: causal power, epistemic access, self-control, and fitting intentions. In other words, an effective oversight person has to have the means to intervene to mitigate risks (causal power), needs enough knowledge to recognize and mitigate them (epistemic access), needs to be in charge of their own doing (e.g., they are not fatigued or drunk) (self-control), and needs to be motivated to do their job properly (fitting intentions). 
While these requirements for the effectiveness of human oversight need to be supported by its general sociotechnical conditions~\cite{sterz_quest_2024,langer_complexities_2025, laux_institutionalised_2023}, in this paper, we are particularly interested in how the \emph{user interface} can support effective human oversight of AI systems.

In this regard, research has mostly been done in the broader area of human-AI decision-making,
where research has focused on providing the user with a sufficient understanding of the AI system.
In particular, the field of XAI (see \citet{ali_explainable_2023, longo_explainable_2024, Langer_2021} for reviews of XAI methods for AI-assisted decision-making) has made great efforts to enhance human understanding of the AI~\cite{miller_explanation_2019, hoffman_metrics_2019, wang_designing_2019} to improve the joint human-AI decision-making performance. Furthermore, providing explanations to the user can support their learning process~\cite{gajos_people_2022, bucinca_contrastive_2025}.
However, AI-based automation can also lead to unintended negative effects on performance when humans are the final decision makers, such as over-reliance on AI, decreases in situational awareness, or de-skilling~\cite{endsley_ironies_2023}. Therefore, recent work has moved beyond automation to explore more active roles for humans in the decision-making process~\cite{danry_dont_2023, chiang_enhancing_2024,ma_towards_2025,reicherts_ai_2025,miller_explainable_2023,gomez_human-ai_2025}, as well as different forms of AI support at various stages of the decision-making process~\cite{ma_beyond_2024,zhang_beyond_2024}.  


Instead of decreasing automation, the goal of this paper is to explore how we can utilize the technical advances made in the field of AI decision-making systems while ensuring that humans can perform effective oversight of these systems to mitigate critical failures or discrimination. 
Beyond AI, there is large body of work on human oversight of automation, where a main concern is the trade-off between automation bias and doing everything themselves~\cite{parasuraman_complacency_2010}. The higher the level of automation of a system, the more supervisory the humans' role has to be~\cite{parasuraman_model_2000}. To support the user in this role, empirical research has focused primarily on cognitive aspects, such as maintaining situational awareness \cite{cummings2006automation, sheridan2019individual}, or on technological features that contribute to oversight performance \cite{endsley2003designing}. 


However, effective oversight requires not only understanding (``epistemic access'') and control (``causal power''), but also sustained motivation and engagement: oversight personnel must be willing and able to perform their role diligently and attentively (i.e., maintain ``fitting intentions''~\cite{sterz_quest_2024}). This points to an overlooked but critical dimension of AI oversight: its work design. 





\subsection{Psychological Processes at Work: Individual Needs and Motivation}
\label{sec:related-work-work-design}
In terms of a human-centered perspective on work tasks more broadly --- beyond human oversight of AI --- research has long emphasized that individuals’ motivation, performance, and satisfaction are deeply conditioned by the extent to which their psychological needs are fulfilled. Self-Determination Theory (SDT) provides a well-established theory for understanding the sources of human motivation \cite{ryan_self-determination_2018, ryan_self-determination_2000, ballou2022}. According to SDT, human motivation is driven by the satisfaction of three fundamental psychological needs: autonomy, competence, and relatedness. When these needs are met, individuals are more likely to be intrinsically motivated to engage in their work tasks \cite{van_den_broeck_review_2016, cerasoli_performance_2016, stanley_meta-analytic_2021,tang_systematic_2020}. 


Beyond SDT, the work design literature has emphasized how specific work characteristics can fulfill fundamental psychological needs and thus foster motivation, performance, and satisfaction. Work design refers to “the content and organization of one's work tasks, activities, relationships, and responsibilities” \cite[p. 662]{parker2014beyond}. 
Recently, researchers have integrated decades of research on work design into an overarching model called SMART which describes five work characteristics that influence psychological needs and long-term job satisfaction (Stimulating, Mastery, Autonomous, Relational, and Tolerable)~\cite{parker2024smart}. In contrast to SDT, which outlines \textit{that} people have fundamental needs and what the consequences of these needs are, SMART emphasizes more closely \textit{how} work characteristics can enable need satisfaction. 

Although the general work design literature provides extensive insights into how to meet employees’ needs and design for good work, including in sociotechnical systems \cite{parker_grote_2020}, only a few studies have addressed these topics in the context of user interfaces for oversight of AI systems~\cite{faas_give_2024}. 
Yet, work design, psychological need satisfaction, and motivation can be considered highly relevant in the context of oversight of AI \cite{sterz_quest_2024}.
In this paper, we thus draw on the work design literature, specifically the SMART model~\cite{parker2024smart} to formulate specific design considerations for the design of oversight interfaces that mediate the structure, content, and organization of oversight work to support both effective and motivational human oversight.

\subsection{Human-centered AI}
At a broader level, the core ideas behind the importance of work characteristics and their contribution to motivation and satisfaction also connect to the field of human-centered AI, which emphasizes designing AI systems that respect and support human values, needs, and capacities \cite{shneiderman2020human}. 
Within HCI, psychological processes, particularly human motivation and autonomy, have long been recognized as central to effective interface design~\cite{horvitz_principles_1999,bellotti_intelligibility_2001, friedman_value-sensitive_1996}. Beyond informing design guidelines, recent research has also examined how specific interface features influence these psychological factors~\cite{bennett_how_2023, bennett_beyond_2024, bucinca_optimizing_2024}. With the advancements of AI technologies and the growing interactions between humans and AI-based systems, the fields of human-AI interaction~\cite{raees_explainable_2024, yang_re-examining_2020, xu_transitioning_2023}, human-centered AI~\cite{shneiderman_human-centered_2020, ozmen_garibay_six_2023, capel_what_2023}, and human-centered XAI~\cite{liao_human-centered_2022, ehsan_human-centered_2022, ehsan_charting_2023} emerged. Central to these fields is the recognition that users’ needs, values, and psychological well-being must guide the design process, leading to design frameworks and guidelines that foreground these aspects~\cite[see][for an overview]{zhao_human-ai_2025} and to a growing emphasis on participatory design approaches for human-AI interaction~\cite{delgado_participatory_2023}. For example, \citet{liao_questioning_2020} collaborated with UX and design practitioners to create an XAI question bank that captures user needs for explainability and guides designers in addressing them. Similarly, \citet{weitz_explaining_2024} explored user requirements in social assessment tasks through a co-design workshop with unemployment consultants, while \citet{kim_help_2023} examined end-users’ needs and perceptions when using a real-world bird identification app. Together, these studies emphasize that understanding stakeholder needs is fundamental to interface design.

Therefore, in our work we opted to conduct co-design workshops with domain experts to directly elicit their perspectives, needs, and priorities when overseeing AI systems. This participatory approach allowed us to ground our design considerations not only in the broader work design literature but also in the experiences of those who are supposed to perform the oversight work, ensuring that the resulting interface concepts not only enhance effectiveness but also support motivation, autonomy, and psychological well-being, core goals of human-centered AI. In doing so, we extend more general design guidelines for human-AI interaction~\cite{apple_human_nodate, google_google_2019, amershi_guidelines_2019} by addressing the specific psychological and motivational challenges of human oversight roles.

%% file: content/method.tex
\section{Method}
The goal of our research is to identify \textit{which aspects we should consider when designing oversight interfaces to help domain experts be effective, motivated and engaged in overseeing AI systems.} 
To this end, we adopted a participatory design approach and complemented it with theoretical insights from work design to interpret and structure our empirical findings into a general design consideration framework. 

We first conducted four co-design workshops, spanning two sessions each, with domain experts from different backgrounds (i.e., computer science, psychology) which we describe in \autoref{sec: oversight task}. With these workshops, we aimed to understand how domain experts approach and experience the oversight of an AI system that automates a task they previously did themselves, what requirements and needs they have for performing this oversight effectively, and how these could be addressed by a user interface.
During the workshop, we first tasked participants to oversee an AI-based grading system for student tests and reflect on their own approaches and experiences (\autoref{sec: appr+exp}). Together, we then discussed their needs and requirements for an effective and motivational oversight interface.In a second session, participants collaboratively prototyped concrete interface designs, critically engaging with these requirements to explore how they could be addressed in practice. (\autoref{sec: reqs}). By actively involving domain experts, we ensured that their perspectives and expertise directly shaped the identification of needs and the design of potential interface solutions.

In a second step, we integrated the empirical findings from our workshop 
with the SMART model of work design  \cite{parker2024smart}, which we introduced in \autoref{sec:related-work-work-design}. In \autoref{sec: smart}, we reflect on how our empirical findings on user requirements and interface prototypes relate to each of the five higher-order work characteristics proposed by SMART (Stimulating, Mastery, Autonomous, Relational, and Tolerable work characteristics) and the psychological processes they condition, and how the theory of SMART can further inform the design of oversight interfaces. By discussing our empirical findings within the scope of each of SMART's work characteristics, we identify a set of specific \emph{design considerations} for developing oversight interfaces.

\section{Co-Design Workshop}
\label{sec: oversight task}
\begin{figure}[ht]
    \centering
    \includegraphics[width=1\linewidth]{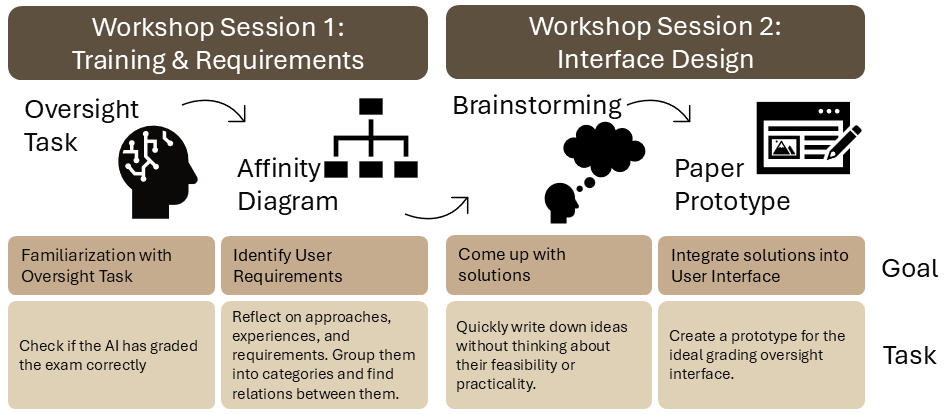}
    \caption{Overview of the workshop design. We performed two sessions with each group, which both consisted of two tasks with different goals.}
    \label{fig: procedure}
    \Description{
    This figure shows an overview of the workshop design. The workshop consisted of two sessions. In the first session, participants had to perform an oversight task to familiarize themselves with the task. The oversight task was to check if the AI accurately graded the student exams. After performing the oversight task, participants had to create an affinity diagram, in which they had to reflect on approaches, experiences, and requirements. These were then grouped into categories, and relations between them were discussed. In the second workshop session, participants first brainstormed solutions, by quickly writing down ideas without thinking about their feasibility or practicality. Lastly, participants created a paper prototype for the ideal grading oversight interface by integrating their solutions into one user interface.
    }
\end{figure}
We performed four co-design workshops with domain experts from different backgrounds (i.e., computer science, psychology).
Each workshop consisted of two two-hour sessions that aimed at different goals (see \autoref{fig: procedure}).
In the first session, participants conducted  the oversight task themselves, and then individually reflected on their approaches and experiences (supported by guiding questions, see \autoref{app:reflection}). With the whole group they then created an affinity diagram guided by the workshop moderator (see \autoref{app:affinity_diagram}) to identify their requirements and needs for support during the oversight process.
In the second session, participants brainstormed and selected solutions for the user requirements they identified in the first session, first individually, then together. The groups then created a paper mock-up of their ideal grading oversight interface. 
Detailed instructions for every part of the workshop are provided in \autoref{app:instructions}.

The oversight task was to oversee an autonomous system that grades student tests. Therefore, participants received a large number of tests (i.e., two exercises with 41 answers each) and had 20 minutes to check if the AI graded them correctly. They received a printed-out sample solution for the exercises, and the students' answers, including the points given by the AI (see \autoref{app: task} for an example task). For the task, we used real questions from the ‘Programming 1’ and ‘Work and Organizational Psychology’ lectures at the authors' home institutions. In the first case, student answers to exercise questions were generated by the OpenAI o4-mini model. In the second case, responses to the exam questions were drawn as random samples from anonymized student exams. The points were assigned using the OpenAI o4-mini model, which was prompted with the grading scheme. See \autoref{app:prompts} for detailed information about how the prompts were structured. Participants were told that due to the examination regulations of the university, a human oversight person was required to check if the autonomous system performed valid point assignments and to identify possible errors. They were instructed to familiarize themselves with the task and think about different approaches they would use if they had to perform this task in the future. 
For detailed instructions, see \autoref{app:task_instruction}.  The workshop was approved by the first author’s institution’s ethical review board.

\subsection{Participants \& Collected Data}
The study was performed in-person with four groups of three participants (see~\autoref{tab:participants}); i.e., \textit{N}=12 (4~Male, 8~Female, age:~20-32, \textit{M}=25.5, \textit{SD}=4.72). The participants were all experts in the task, meaning they had prior grading experience and expertise in the respective domain (i.e., psychology or computer science). We decided to perform the task with computer scientists and psychologists to collect a broad range of ideas, approaches, and user requirements, and to make our findings more generalizable. In a pre-workshop questionnaire, participants indicated that they have not used AI support for grading student exams or assignments before (1-2~on a 5-point Likert scale, \textit{M}=1, \textit{SD}=0.39). Participants were recruited by reaching out to researchers in psychology and former tutors of the 'Programming 1' lecture at the authors' institutions.


\begin{table*}[ht]
    \centering
    \begin{tabular}{cccc}
    Group ID &  Age & Expertise &  Material  \\
    \toprule
    CS1 & 21-23 & CS Bachelor and Master Students & Programming 1 Test \\
    CS2 &  20-22 & CS Bachelor Students & Programming 1 Test \\
    Psy1 & 30-32& Psychology PhD Students and Postdoc& W\&O Psychology Exam \\
    Psy2 & 28-31 & Psychology PhD Students & W\&O Psychology Exam \\
    \end{tabular}
    \caption{Demographics of the different workshop groups}
    \label{tab:participants}
\end{table*}

The pre-workshop questionnaire included items on participant demographics, their grading experience, and attitude towards AI support tools.
Participants created an affinity diagram of their approaches, experiences, and user requirements when performing the oversight task, collected ideas about how a user interface could support them in this task, and created a paper mock-up of a user interface for the oversight task. 
The participants' answers to the pre-workshop questionnaire, the affinity diagrams, the ideas collected for a user interface, and pictures of the final paper mock-ups are provided in the supplementary materials. All workshops were video- and audio-recorded. They were performed in English (CS2) or German (CS1, Psy1, Psy2). For this paper, participants' quotes were translated and shortened for clarity.

\subsection{Data Analysis Approach}
To analyze the data collected during the workshops, we employed reflexive thematic analysis~\cite{braun_using_2006,braun_reflecting_2019}. 
It supports inductive, flexible theme development while allowing the researchers’ perspectives and expertise to play a meaningful role in the analysis.
We did not transcribe the video and audio recordings. Instead, coding was performed directly on the video materials to preserve contextual and nonverbal cues (e.g., the spatial layout of affinity diagrams, gestures, and emphasis in group discussions).
Three researchers (all co-authors of this paper) conducted the analysis. Coders 1 and 2 each analyzed the workshop(s) they had facilitated. A third coder, who had not facilitated any session, coded all workshops independently to enhance interpretive diversity and support analytical rigor. All authors then participated in identifying and refining themes from the codes. 
The qualitative analysis proceeded through the following steps:
\begin{enumerate}[labelindent=\parindent,leftmargin=*,label={\arabic*.}]
    \item \textit{Familiarization.} Each coder reviewed their assigned workshop materials fully to immerse themselves in the data.
    \item \textit{Initial Coding.} The coders developed descriptive and interpretive codes that closely reflected the participants’ own language and captured their relevant thoughts. 
    \item \textit{Theme Development.} The codes were grouped into broader conceptual categories, i.e., themes, to identify recurring patterns and points of convergence related to the research questions.
    \item \textit{Theme Refinement.} Themes were iteratively reviewed to ensure internal coherence and clear distinction. During this process, some themes were merged, split, or redefined to more accurately represent the underlying data. 
    \item \textit{Theme Naming.} Each theme was given a descriptive name that reflected its central concept and function within the broader thematic structure and the researchers’ higher-level interpretation of recurring ideas in the data. 
    \item \textit{Relationship Construction.} Coders examined the relations among themes within and across research questions. This included identifying conceptual overlaps, as well as inferred cause-effect relationships.
\end{enumerate}
These stages were informed by both the researchers’ close engagement with the data and their interpretive synthesis, meaning the process of drawing meaning from patterns across the data based on analytical judgment, contextual understanding, and iterative reflection. Importantly, this process was also guided by the researchers’ prior empirical and theoretical knowledge, particularly from the previously discussed literature.
Reflexive thematic analysis explicitly allows for such theoretically-informed interpretation and positions researcher subjectivity as a resource rather than a limitation. Through our analysis approach with three independent researchers from diverse backgrounds (Computer Science \& Philosophy (Coder 1) and Work \& Organizational Psychology (Coder 2,3)), we ensured interpretive diversity and analytical rigor.
Finally, the coding team synthesized their results collaboratively. Through discussion, the coders compared and reconciled interpretations, refined the thematic structure, and jointly confirmed the relationships between themes. The final thematic framework is presented in \autoref{fig: findings}.  
In the following, we first present results on the approaches and experiences of the domain experts (\autoref{sec: appr+exp}) and then on the identified user requirements and their implementation in UI prototypes (\autoref{sec: reqs}).

%% file: content/results.tex
\label{sec:results}
\begin{figure}[ht]
    \centering
    \includegraphics[width=1\linewidth]{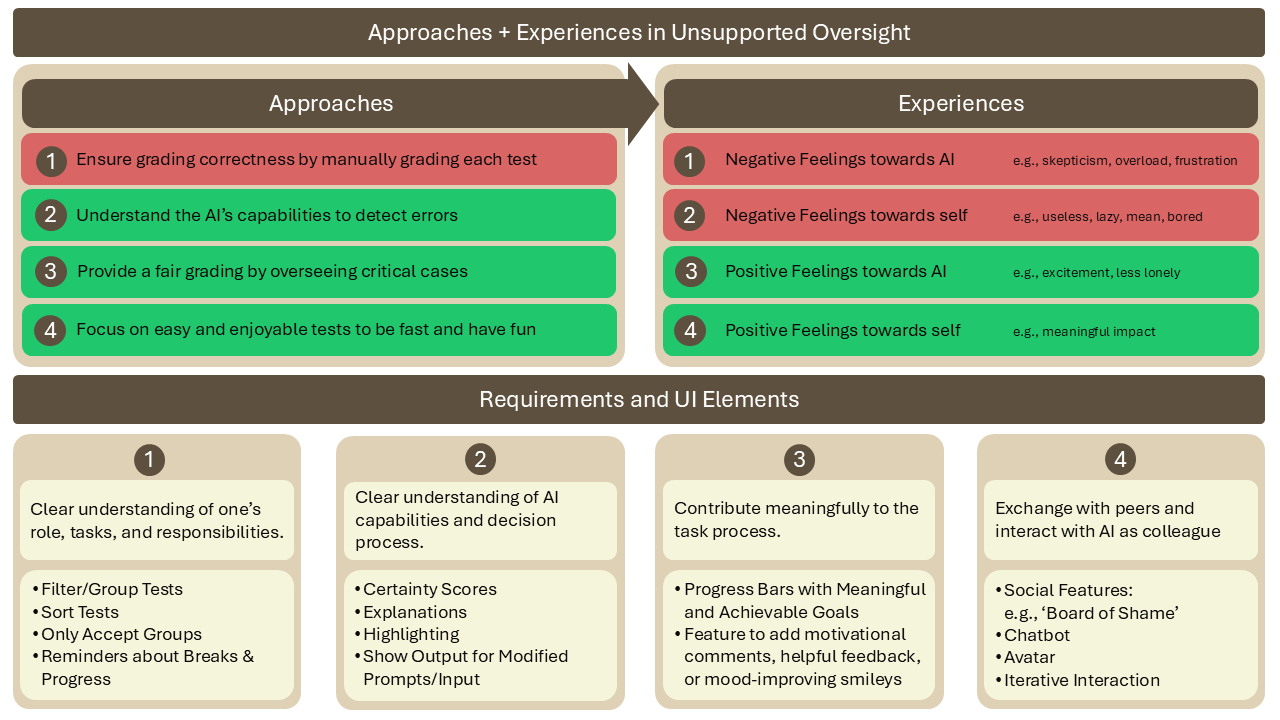}
    \caption{Overview of the main themes identified in our thematic analysis.}
    \label{fig: findings}
    \Description{
    This figure provides an overview of the main themes that were identified in our thematic analysis and which are discussed in the results sections.
    }
\end{figure}

\subsection{Results: Initial Approaches and Experiences of Domain experts}
\label{sec: appr+exp}
The most prevalent finding across all groups was that participants were drawn to \textit{perform the AI's grading task} instead of fulfilling their role as an oversight person. 
This approach undermined efficiency gains and led to  mainly \textit{negative feelings} when performing the task. Approaches that participants perceived as more positive included detecting unjustly awarded points by understanding the AI’s strengths and weaknesses, ensuring fair grading of borderline cases, and focusing on tests that were easy and enjoyable to grade.
For each of the approaches, we discuss the underlying goal and the strategy the participants adopted to achieve it below, as well as the positive and negative experiences participants reported with these approaches.


\subsubsection{Approaches}
\label{sec: approaches}

\begin{enumerate}[labelindent=\parindent,leftmargin=*,label={Approach 1.}]
    \item Ensure grading correctness by manually grading each test.
\end{enumerate}
Most participants initially did not approach the task as an oversight task, but as a grading task, which was characterized by the participants' goal to check all of the students' tests (CS1, CS2, Psy1, Psy2) by grading them manually and comparing their grade to the AI's grade (CS1, CS2, Psy1, Psy2). Reflecting on this initial approach, participants pointed out that their goal was not achievable and the strategy, therefore, was not efficient (CS1, CS2, Psy1, Psy2; Psy 1, P1: „I didn’t really find it efficient. You still had to read everything, you still had to assess everything and then compare that with your own grading“).
 
\begin{enumerate}[labelindent=\parindent,leftmargin=*,label={Approach 2.}]
    \item Detect unjustly awarded points by understanding the AI's strengths and weaknesses.
\end{enumerate}
After reflecting on their initial approach, participants described that the goal to check only a subset of the tests was more achievable than aiming for all tests (CS1, CS2, Psy1, Psy2) and that strategies that included skimming the answers much more superficially, as compared to when doing the grading manually, are more efficient (Psy1). Therefore, it was proposed to aim to find unjustly awarded points (CS1, Psy1) by understanding the strengths and weaknesses of the AI system and focus only on a subset of tests, which they judged especially important for their task (CS1, CS2, Psy1, Psy2). Participants deemed extreme cases important, such as tests with zero or full points or unique answers, since they expected the AI to perform worse in these cases (CS1, CS2, Psy1, Psy2).

\begin{enumerate}[labelindent=\parindent,leftmargin=*,label={Approach 3.}]
    \item Provide fair grading by overseeing critical cases.
\end{enumerate}
Another goal of the participants was to provide fair grades (CS1, CS2, Psy1) by focusing on subsets of tests, which they judged especially important (CS1, CS2, Psy1, Psy2). This resulted in the approach to identify critical cases in which the AI assigned too few points, causing the student to fail the test (CS1). 

\begin{enumerate}[labelindent=\parindent,leftmargin=*,label={Approach 4.}]
    \item Perform the task fast and have fun by overseeing tests that are easy and enjoyable to grade.
\end{enumerate}
Lastly, participants discussed the goal of performing the task fast and having fun (CS1, CS2, Psy1, Psy2). This goal was pursued by focusing on the tasks that were easiest (CS1, CS2) or most fun to check (CS2).


\subsubsection{Experiences}
\label{sec: experiences}
\begin{enumerate}[labelindent=\parindent,leftmargin=*,label={Experience 1.}]
    \item Negative feelings towards AI.
\end{enumerate} 
Most negative experiences related to their feeling that human-AI collaboration was not beneficial for this task. Participants felt skeptical of AI (Psy1), unsupported (CS1, CS2, Psy1, Psy2), or like doing extra work (CS1, Psy1, Psy2). They felt negative overall (CS1, Psy1), frustrated (CS1, Psy1), stressed (Psy2), confused (Psy1), uncertain (Psy1), and biased (CS1, CS2, Psy1). One group (CS1) linked their negative experience to their perceived responsibility for possible errors, since they felt obligated towards the students that humans should assess their tests: "Somebody should be responsible for the grade, and we can't say it is the AI" (CS1, Participant 2).
\begin{enumerate}[labelindent=\parindent,leftmargin=*,label={Experience 2.}]
    \item Negative feelings towards self.
\end{enumerate} 
Performing the oversight task also caused negative feelings about themselves. One participant (CS1, P1) described that they focused on tests, where the AI gave full points, which seemed most efficient, but made them feel bad about themselves: "In the end, if you decrease points from people, you think, if I did not change it, they would have more points and I screwed them up. That's the feeling that resonates."(CS1, P1). Other participants felt useless performing the task (CS2), which made them feel lazy (CS1), bored (CS2, Psy1), or tired (CS2, Psy1).
\begin{enumerate}[labelindent=\parindent,leftmargin=*,label={Experience 3.}]
    \item Positive feelings towards AI support.
\end{enumerate}
Despite the rather negative experiences, some participants felt hopeful or excited based on the prospect of saving time and energy (Psy1). One group (Psy1) reflected that the valence of their experiences depended on the specific cases they were overseeing. For example, some participants judged the AI's grading as very good for one of the two exercises they received, but did not understand its recommendation for the other (Psy1, P2: "There is definitely satisfaction [with the system] but kind of limited to specific questions").
\begin{enumerate}[labelindent=\parindent,leftmargin=*,label={Experience 4.}]
    \item Positive feelings towards self.
\end{enumerate}
Furthermore, one participant felt less lonely when performing the task and safer in their decision-making, because they could share the responsibility for the outcome (Psy2), and another approached the task intending to have a meaningful impact (CS1, P2). Aiming to prevent students from failing the test due to an AI error, they had a more positive experience than the other participants in their group: "Too many points do not hurt the student, but rather if the grading is done too strictly; I think it is somehow wrong [...] to always decrease their gradings"(CS1, P2). 

\subsection{Results: User Requirements and Supporting UI Elements}
\label{sec: reqs}
Reflecting on the oversight task, the participants described different user requirements for a good oversight interface. We found four main themes when discussing their user requirements and needs: role understanding, AI understanding, meaningfulness, and relational aspects. Participants also came up with UI elements, that support the interface to meet their requirements.

\subsubsection{Role Understanding: Clear understanding of one's role, the associated responsibilities, and tasks.}
\label{sec:req-role}

Participants desired a clear understanding of their role, which should be reflected in the user interface's affordances. When performing the oversight task, they wished for a clear understanding of their responsibilities (CS1, Psy1, Psy2), since the task itself did not clarify the responsibility distribution, resulting in varying perceptions. Furthermore, they wanted the interface to clearly represent what their task is to avoid misconceptions based on prior experiences: (Psy1, Psy2; Psy2, P2: "I was aware that I should oversee the AI, but because of the formatting of the answers, I reverted to old habits of grading exams as I did it in the past").

\begin{figure}[ht]
    \centering
    \begin{subfigure}{0.26\linewidth}
        \centering
        \includegraphics[width=\linewidth]{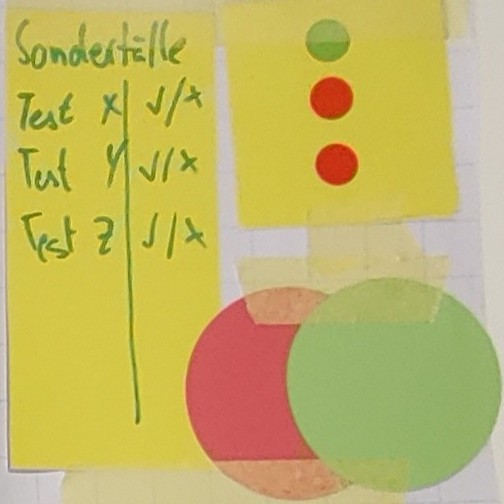}
        \caption{Grouping of special cases with progress bar - CS1}
    \end{subfigure}
    \begin{subfigure}{0.26\linewidth}
        \centering
        \includegraphics[width=\linewidth]{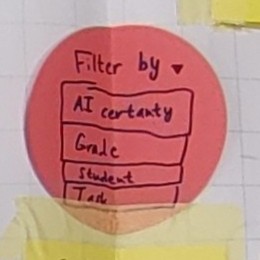}
        \caption{Function to filter tests according to different criteria - CS2}
    \end{subfigure}
    \begin{subfigure}{0.26\linewidth}
        \centering
        \includegraphics[width=\linewidth]{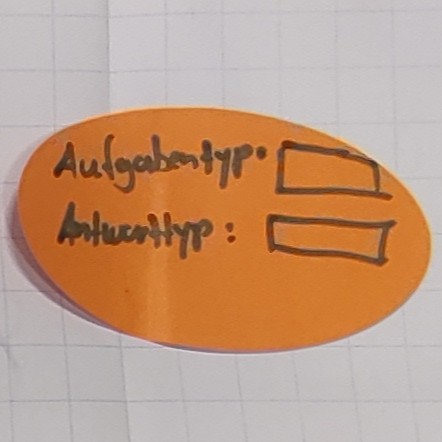}
        \caption{Selection for different kinds of tasks or answers - Psy2}
    \end{subfigure}
    \caption{UI Elements that participants included to filter and group the student answers.}
    \label{fig: filterAndGroup}
    \Description{
    This figure shows UI elements from three workshop groups that included filtering and grouping the students' answers. Group CS1 had one part that showed tests that were categorized as special cases by the AI. Group CS2 included a filtering function, where they could filter the tests according to, e.g., AI certainty, grades, student, or task. Group Psy2 included two drop-down menus to select a certain exercise type or answer type one wants to be displayed.
    }
\end{figure}

Therefore, two groups actively decided not to include an accept/reject decision for every single test, but only a button to finish the task (CS2, Psy2), and one group separated the task of detecting errors from fixing them (Psy2).
All workshop groups tried to focus on subsets of tests that are important to oversee by grouping or sorting the tests according to different criteria (CS1, CS2, Psy1, Psy2) (see \autoref{fig: filterAndGroup}). 
Additionally, participants aimed to remind users about their oversight role and support them in not grading too many tests themselves by including reminders about taking breaks (Psy2) or finishing the task after a certain time (CS1, CS2). 


\subsubsection{AI Understanding: Clear understanding of AI capabilities and decision process.}
\label{sec:req-AI-understanding}
\begin{figure}[ht]
    \centering
    \begin{subfigure}{0.3\linewidth}
        \centering
        \includegraphics[height=0.41\linewidth]{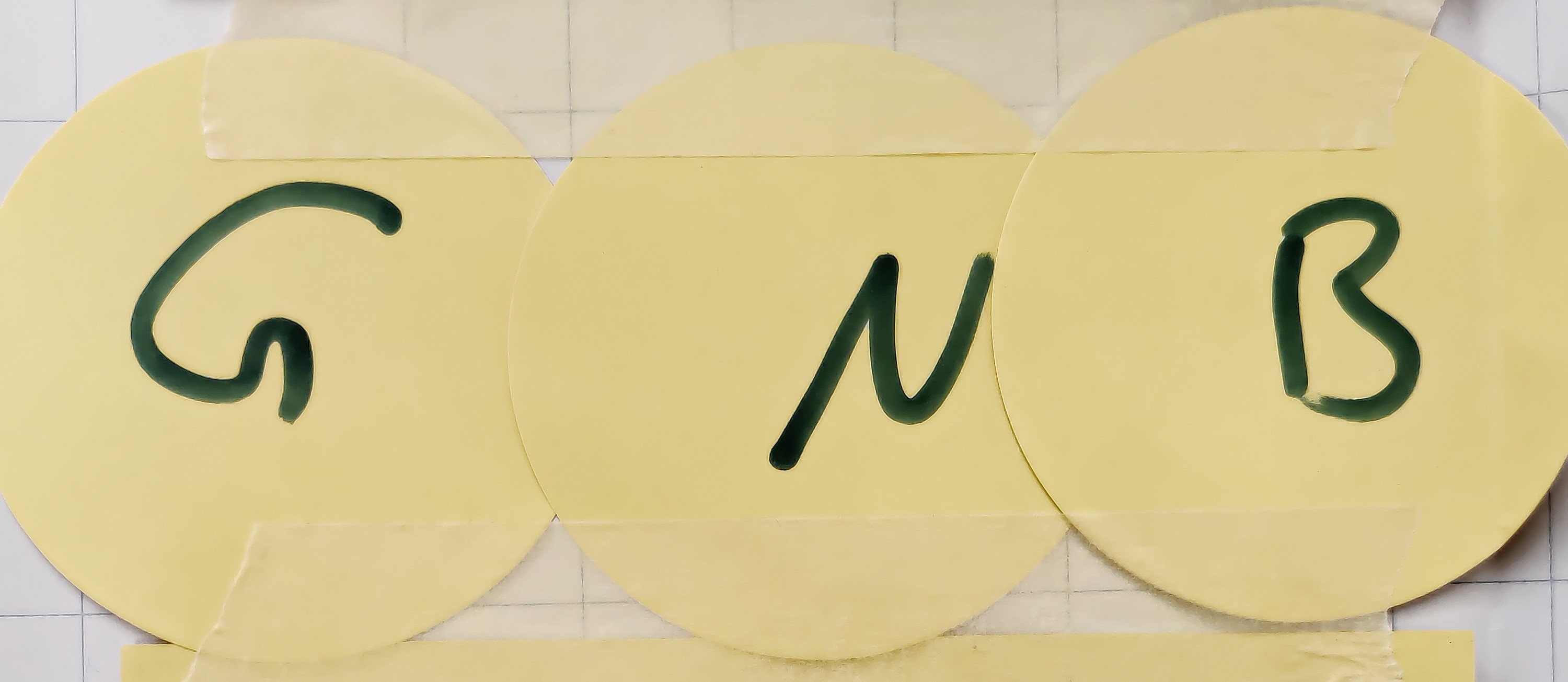}
        \caption{Prompt to grade with 'good',\newline 'neutral', or 'bad' sentiment - CS1}
    \end{subfigure}
    \begin{subfigure}{0.25\linewidth}
        \centering
        \includegraphics[height=0.49\linewidth]{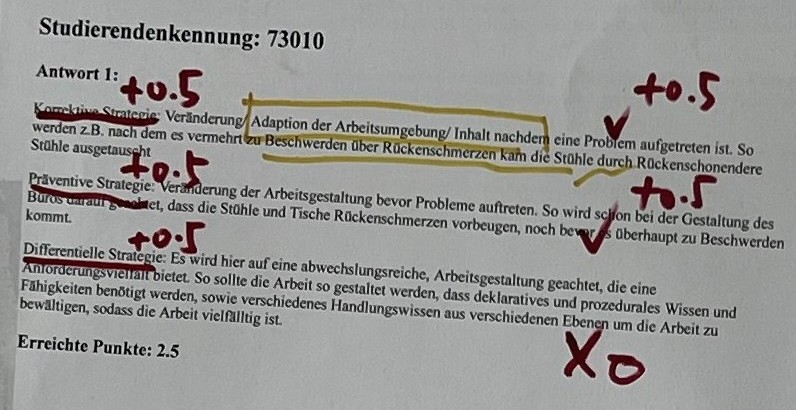}
        \caption{AI highlights reasoning\newline for points - Psy1}
    \end{subfigure}
    \begin{subfigure}{0.35\linewidth}
        \centering
        \includegraphics[height=0.35\linewidth]{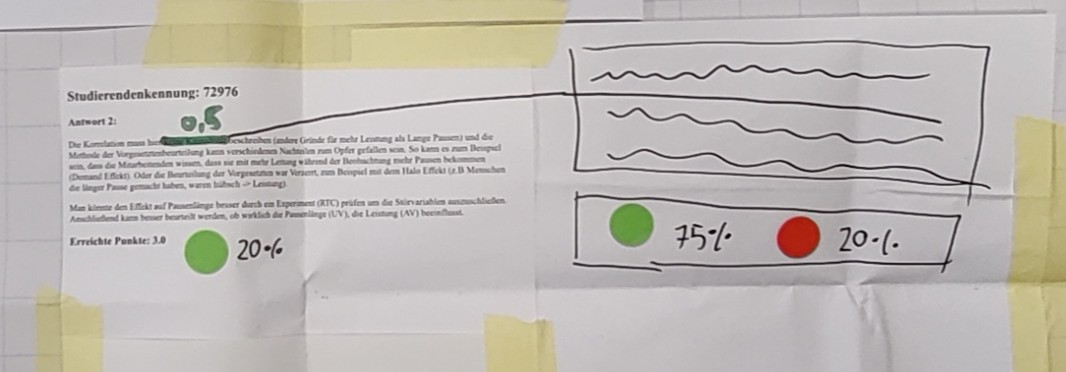}
        \caption{Confidence scores and explanations\newline for how points were given - Psy2}
    \end{subfigure}
    \caption{UI Elements that participants included to better understand the AI capabilities and its decision process.}
    \label{fig: AICapabilities}
    \Description{
    This figure shows UI elements that are aimed to support the oversight person in understanding the AI's capabilities and decision-making process. Group CS1 displayed a confidence score and provided the option to modify the AI prompt and let it grade the test with 'good', 'neutral', or 'bad' sentiment. Group Psy1 included a confidence score for the decision, as well as highlighting of parts in the answer, which gave points. Group Psy2 included confidence/certainty scores for the decision, as well as highlighting of parts in the answer, which gave points and explanations why this part of the answer is worth points.
    }
\end{figure}

Participants expressed their desire to understand the AI's capabilities and its decision process and requested system features to support them (see~\autoref{fig: AICapabilities}) (CS1, CS2, Psy1, Psy2). Participants wished for information about the AI's strengths and weaknesses, such as certainty/confidence scores (CS1, CS2, Psy2), explanations (CS2), or accuracies (Psy2). Further, they desired to understand the AI's reasoning and decision-making process (Psy2, P2: "It should be clear --- the steps behind --- how the AI came to the decision"; P1: "We mentioned that a lot: for us transparency is very important"). The UI should show (parts of) the process that led to the decision by highlighting decisive parts of the answer (CS1, Psy1, Psy2), providing a general grading policy (CS1), assessing the answer based on various prompts (CS1), or providing an improved student answer (CS1). 
Interestingly, they hypothesized that their little understanding of the AI also decreased their role understanding: "I think the difficulty of focusing on overseeing the system is that you don't have much information about the system. Otherwise, this shift would have been easier." (Psy2, P33). 

\subsubsection{Meaningfulness: Contribute meaningfully to the task process.}
\label{sec:rec-meaningfulness}
Participants wanted to perform a meaningful task. In group CS1, participants found that pursuing a meaningful goal (i.e., detecting correct features of answers that the AI missed) made them feel satisfied.
They saw opportunities for meaningful additions to AI grading that would support students, such as motivational comments, helpful feedback, or mood-enhancing smileys (CS1, CS2, Psy2). 
However, meaningfulness for them was (at least partially) constituted by their role understanding, performance, and control over the AI and outcomes.
In their interface prototypes, meaningful goals were mainly apparent through progress bars, which would only benefit them if they aimed for meaningful goals (CS1, CS2, Psy1, Psy2). Therefore, two groups included progress bars towards one or multiple small sub-goals (CS1, CS2), and one group (Psy2) removed the progress bar again. But also providing confidence levels for the grading, they would "start with the things that AI has the lowest confidence in—where they need me most" (Psy1, P2).


\subsubsection{Relational Aspects: Exchange with peers and interact with AI as colleague.}
\begin{figure}[ht]
    \centering
    \begin{subfigure}{0.3\linewidth}
        \centering
        \includegraphics[width=.65\linewidth]{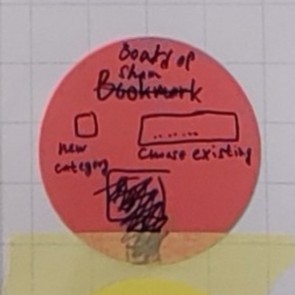}
        \caption{Board of Shame - CS2}
        \label{fig: bookOfShame}
    \end{subfigure}
     \begin{subfigure}{0.3\linewidth}
        \centering
         \includegraphics[width=.7\linewidth]{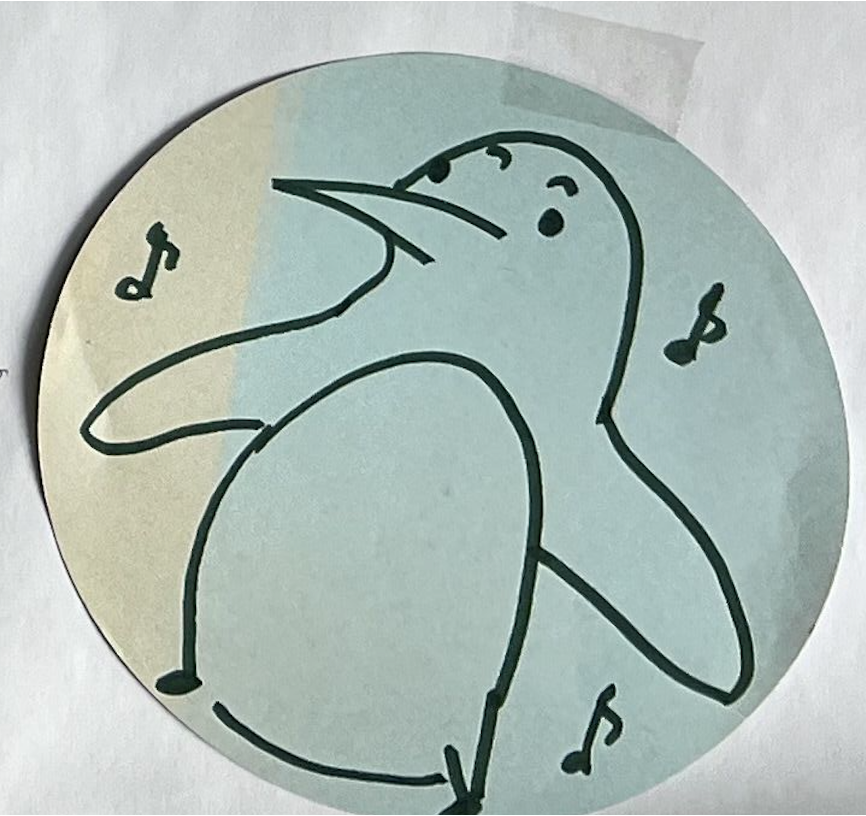}
    \caption{Avatar - Psy1}
        \label{fig: avatar}
        \end{subfigure}
    \begin{subfigure}{0.3\linewidth}
        \centering
        \includegraphics[width=\linewidth]{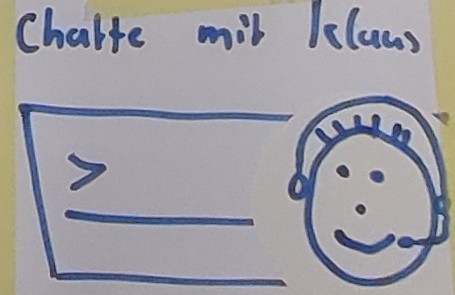}
        \caption{Chatbot - Psy2}
        \label{fig: chatbot}
    \end{subfigure}
    \caption{UI Elements that participants included in their prototypes to meet their relational requirements.}
    \label{fig: relAsp}
    \Description{
    This figure shows UI elements that participants included in their prototypes to meet their relational requirements. Group CS2 included a 'Board of Shame' which can be used to save weird answers and exchange with other students. Group Psy1 included an avatar that looked like a penguin, which sometimes started dancing. Group Psy2 included a chatbot with the name Klaus and a human-like face.
    }
\end{figure}

A common theme that occurred in all workshop groups was the desire for social interaction when performing the oversight task. This social interaction could take different forms: with other oversight people, the students, or the AI. In their prototypes, participants included features that supported them to meet their relational needs (see \autoref{fig: relAsp}). 

Participants described the desire to connect with other oversight people by exchanging about different student answers they encountered (CS1, CS2). One group (CS2) added a 'Board of Shame'(\autoref{fig: bookOfShame}) to the interface where they could share funny or creative student answers with other tutors: "If the button was not called 'Bookmark' but 'Board of Shame', [...] I would for sure bookmark every weird edge-case I encountered. That would be more fun." (CS2, P3) [...] "that would be extremely motivating." (CS2, P1).
Furthermore, social interactions were important for deciding which information to display in the interface (CS1).

Participants wanted to have human-like interactions with AI to feel connected (Psy2) or experience the task as more enjoyable and fun (Psy1, Psy2). Group Psy2 added a chatbot with a human-like name and appearance, while Group Psy1 used the likeness of a penguin, which dances or smiles when good student answers were displayed on the screen. (\autoref{fig: chatbot}). They wanted the AI to appear as a colleague with which they can communicate and discuss to understand its recommendations better  
, improve the grading, and share responsibility (CS1, Psy1, Psy2). Notably, group Psy1 stressed how this relational need and its fulfillment were prone to inter-individual differences. 

%% file: content/discussion.tex
\section{Theoretical Integration and Design Considerations}
\label{sec: smart}


In the following, we reflect on how our empirical findings on user requirements and interface prototypes relate to each of the five higher-order work characteristics proposed by SMART (Stimulating, Mastery, Autonomous, Relational, and Tolerable work characteristics) and the mediating psychological states (e.g., work meaningfulness, fulfillment of relational needs) that ultimately shape job satisfaction. Although SMART primarily addresses work design at the job level, these characteristics also apply to user interfaces and technologies through which work is performed, influencing the same psychological processes. Considering these characteristics at the interface design level could thus help to design user interfaces that are not only functional but contribute to job satisfaction and long-term performance~\cite{parker2024smart}. 
By discussing our empirical findings within the scope of each of SMART's work characteristics, we identify a set of \emph{design considerations} for developing oversight interfaces, summarized in Table~\ref{tab:design-guidelines}. These design considerations are not prescriptive guidelines but rather points for reflection, acknowledging that enhancing one characteristic may trade off with another or alter the oversight task. \autoref{tab:design-guidelines} further illustrates how participants expressed each consideration in their interface prototypes.

\begin{table}[!ht]
\footnotesize
\sffamily
\centering

\begin{tabular}{p{0.35\linewidth} p{0.4\linewidth} p{0.18\linewidth}}
\toprule

\textbf{ Design Consideration for Oversight Interfaces} & \textbf{ Example from Co-Design Workshop and Grading Task} & \textbf{ Psychological Processes} \\
\midrule
\midrule

\multicolumn{3}{l}{\textbf{Stimulating Interface Characteristics\smallskip}} \\

\emph{Skill Variety.} \newline Consider how the interface can encourage variety in the skills applied during oversight. 
& The interface offers the user a varied set of tests to check, including different exercise types, answer lengths, or different types of student mistakes, to make the oversight task less monotonous and more stimulating.  \smallskip 
& Challenge Appraisals and Work Meaningfulness \\

\emph{Meaningfulness of oversight.} \newline Consider how the user can contribute to the automated task in a way that is meaningful to them and reflect that in the interface.  
& The interface displays borderline cases in a dedicated area to help oversight personnel ensure that no student unjustly fails the test.
&  \\

\midrule
\multicolumn{3}{l}{\textbf{Mastery Interface Characteristics}\smallskip} \\

\emph{Understanding of AI.} \newline Consider which information about the AI's capabilities and decision process supports the user in their oversight task. 
& For each type of exercise, the interface displays an average certainty value of the AI grading. This could help the user understand the AI’s overall strengths and weaknesses better than showing certainties for each individual test.  \smallskip
& Challenge Appraisals and Activated Negative Affect \\

\emph{Understanding of Oversight Role.} \newline Consider how UI elements influence the user's role perception. \smallskip
& The interface has a "change grading" button but no "accept" button to prevent the user from re-grading each individual test. 
&  \\

\emph{Feedback.} \newline Consider how the user interface can indicate task progress and the impact of interventions on the final output or the AI system. 
& The interface displays statistics and metrics about the grading, such as point distributions. If the oversight person adjusts the grading scheme or lets the AI reevaluate the tests, they immediately see the overall impact of their intervention.
&  \\

\midrule
\multicolumn{3}{l}{\textbf{Autonomous Interface Characteristics}\smallskip} \\

\emph{Timing Autonomy.} \newline Depending on the timing constraints, consider how the interface can give the user autonomy to organize their responsibilities. \smallskip
& The interface offers the user the possibility to "save" individual tests to a dedicated area from which they can be re-graded at any time. This is possible since grading is an oversight task without time-critical interventions. 
& Work Meaningfulness \\

\emph{Method Autonomy.} \newline Depending on the oversight constraints, consider how the interface can give the user autonomy about how they monitor the AI and how they intervene.  
& The interface offers sorting and grouping options to allow the user to choose and prioritize which tests to review.
&  \\

\midrule
\multicolumn{3}{l}{\textbf{Relational Interface Characteristics}\smallskip} \\

\emph{Relationship to peers.} \newline Consider how the interface can support the user in relating to other oversight persons. \smallskip
& The interface provides the option to share interesting answers with other people involved in the oversight, which they might discuss outside the oversight process.
& Work Meaningfulness and Fulfillment of Relational Needs \smallskip\\

\emph{Relationship to affected individuals.} \newline Consider mechanisms that help users to recognize and positively shape the impact of their oversight on others.  \smallskip
& The interface supports allows the user to add motivational comments or constructive feedback to the automated grading.
& \\

\emph{Relationship to AI.} \newline Consider how the appearance and interaction with the overseen AI contribute to the user's relational needs but might affect oversight performance. 
& The interface lets the user exchange with a human-like AI about the decisions, which lets the oversight person connect more easily with them, but may increase bias.
&  \\

\midrule
\multicolumn{3}{l}{\textbf{Tolerable Interface Characteristics}\smallskip} \\

\emph{Role Overload.} \newline Consider how the interface can present the requirements of the oversight task in a manageable way. 
& The interface provides metrics on different exercises and overall grading, which help the user manage their resources efficiently.\smallskip
& Activated Negative Affect \\

\emph{Role Conflicts.} \newline Consider how the interface can clearly separate the tasks and responsibilities of the user from those of the AI. 
& The interface allows users to improve the AI’s grading solely by requesting a reevaluation, which reduces role conflict, as users do not grade tests themselves.
&  \\

\bottomrule
\end{tabular}
\caption{Considerations for designing oversight interfaces, grouped by their mapping to the Work Characteristics proposed by the SMART model of work design~\cite{parker2024smart} and links to psychological processes as proposed by SMART. Each consideration is illustrated by an example from our co-design workshop that demonstrates how participants considered each point in their own prototypes.}
\label{tab:design-guidelines}
\end{table}

\subsection{Stimulating work characteristics} 
\label{sec: stimulating}
Stimulating work is characterized by task variety and the opportunity to draw on different skills and requirements for problem-solving and information processing. When these features are absent, work tends to be experienced as monotonous and demotivating. In our study, oversight resembled the repetitive, low-level tasks that SMART links to reduced stimulation, with participants describing re-checking straightforward AI outputs as boring and tiring. Oversight appeared more engaging when participants focused on ambiguous, extreme, or idiosyncratic cases, which SMART suggests may enhance cognitive challenge and stimulation.

Oversight interfaces should thus contribute to oversight persons being able to use different skills for problem-solving and information processing. 
Participants in our workshop recognized that some tests were more stimulating to check than others and added interface elements to highlight them, for example, by displaying outliers or ambiguous answers in a dedicated space, or by enabling sorting based on AI certainty or received points (see \autoref{fig: filterAndGroup}).

\begin{enumerate}[labelindent=\parindent,leftmargin=*,label={\textbf{C1.}}]
    \item \textbf{Skill Variety.} Consider how the interface can encourage variety in the skills required for oversight.
\end{enumerate}

The pronounced concern among participants to contribute meaningfully to the process (see \autoref{sec:rec-meaningfulness}) and to perform the task well, suggests that the task was not experienced as sufficiently stimulating on its own. According to SMART, insufficient stimulation diminishes perceived meaningfulness, which in turn lowers motivation and satisfaction, as evidenced by the participants' negative experiences (\autoref{sec: experiences}). 

Thus, designers should consider how the user interface can support domain experts in contributing meaningfully by overseeing the automated task. 
Participants saw meaning in preventing students from unjustly failing the exam. Thus, they integrated filtering and sorting features to focus on students who were just below the passing threshold.

\begin{enumerate}[labelindent=\parindent,leftmargin=*,label={\textbf{C2.}}]
    \item \textbf{Meaningfulness of oversight.} Consider how the user can contribute to the automated task in a way that is meaningful to them and reflect that in the interface.
\end{enumerate}

\subsection{Mastery work characteristics} 
\label{sec: mastery}
According to SMART, workers require an understanding of what their tasks are, how these tasks fit into the broader process, and how well they are performing. The salience of this issue was reflected in the themes of role understanding (\autoref{sec:req-role}) and AI understanding (\autoref{sec:req-AI-understanding}).  
Participants were often uncertain whether they should grade or oversee, unclear about their responsibilities, and lacked insight into the AI’s decision-making, which might have helped them better understand their role and contribution.
Uncertainty about their role undermines mastery and fosters negative affect, which aligned with participants’ negative experiences of uselessness and overload (\autoref{sec: experiences}).

Designers should consider UI elements that support understanding and monitoring of AI decisions, clarify the oversight task through intervention mechanisms, and provide feedback on task progress and the impact of human oversight.

In their interface prototypes, participants included features like confidence scores and AI output explanations. One participant suggested displaying the AI’s grading policy rather than individual test results to promote higher-level oversight, expecting this would enhance understanding of both the AI’s capabilities and their own oversight responsibilities.
Some observed that the option to directly modify grades encouraged them to take over the grading task, leading to its removal from the interface. Instead, one group proposed allowing the AI to re-evaluate tests based on additional input or instructions.
Most participants included progress bars, indicating a need for UI support in tracking performance. Additional feedback could highlight the impact of their interventions, for example, metrics showing how many students passed due to their oversight, how many AI errors they identified, or how re-evaluations they triggered changed the grading.

\begin{enumerate}[labelindent=\parindent,leftmargin=*,label={\textbf{C3.}}]
    \item \textbf{Understanding of AI.} Consider which information about the AI's capabilities and decision process supports the user in their oversight task.
\end{enumerate}

\begin{enumerate}[labelindent=\parindent,leftmargin=*,label={\textbf{C4.}}]
    \item \textbf{Understanding of Oversight Role.} Consider how UI elements influence the user's role perception.
\end{enumerate}

\begin{enumerate}[labelindent=\parindent,leftmargin=*,label={\textbf{C5.}}]
    \item \textbf{Feedback.} Consider how the user interface can indicate task progress and the impact of interventions on the final output or the AI system.
\end{enumerate}



\subsection{Autonomous work characteristics} 
\label{sec: autonomous}
According to SMART, autonomy refers to perceived control over timing, methods, and decisions, enabling ownership, which in turn promotes a sense of meaningful work.
Autonomy emerged as a recurring tension in the oversight task. 
Because automated grading is not time-sensitive, participants generally had high timing autonomy and rarely operationalized it. While some expressed a desire to time their actions freely, they chose to limit this autonomy to maintain clearer role boundaries. This raises the question of whether UI support for timing autonomy may be more critical in time-sensitive tasks.
While participants had method autonomy in how to perform oversight, the accompanying ambiguity about decision rights and accountability weakened their sense of control and ownership (see \autoref{sec: mastery}). This helps explain participants’ frustration and doubts about their role’s purpose, suggesting that method autonomy alone is insufficient for meaningful work without clearly defined scope and limits.

These tensions highlight the need for designers to provide users with an appropriate degree of autonomy, clearly communicated through the interface. For work to feel meaningful, users require both autonomy and a clear understanding of their role. Designers should define which freedoms align with the user's role and task, make role boundaries visible, and offer meaningful choices within them. While too many options can overwhelm users, too little autonomy can leave them feeling restricted and ineffective.

One group discussed that oversight involved both monitoring the AI’s grading and correcting errors. To avoid reverting to manual grading, they preferred postponing corrections and felt that limiting their ability to intervene directly would enhance role clarity by reducing reminders of their previous grading role. Importantly, they valued timing autonomy, but omitted it in favor of preserving role clarity.
Participants exercised method autonomy primarily by choosing which subset of student answers to oversee, but also discussed allowing users to choose their level of oversight, such as reviewing individual tests, the AI’s grading policy, or grading metrics. Designers should also consider what types of interventions to support, such as correcting the error themselves, delegating it, or prompting the AI to re-evaluate specific tests.

\begin{enumerate}[labelindent=\parindent,leftmargin=*,label={\textbf{C6.}}]
    \item \textbf{Timing Autonomy.} Depending on the timing constraints, consider how the interface can give the user autonomy to organize their responsibilities..
\end{enumerate}

\begin{enumerate}[labelindent=\parindent,leftmargin=*,label={\textbf{C7.}}]
    \item \textbf{Method Autonomy.} Depending on the oversight constraints, consider how the interface can give the user autonomy about how they monitor the AI and how they intervene.
\end{enumerate}

\subsection{Relational work characteristics} 
\label{sec: relational}
Relational work characteristics support the need for relatedness by enabling social interaction and a sense of contributing to others within a broader context.
Since the oversight task offers limited opportunities for relational fulfillment, participants expressed their desire for exchanges with peers about unusual or noteworthy student answers and connecting with students through meaningful feedback, motivational comments, or small gestures of recognition. From a SMART perspective, these social and beneficiary-oriented contributions support the psychological need for relatedness.

Interface designers should consider how the UI can support users' relational needs, not only through social features, but also by displaying information that facilitates later peer exchange. Additionally, when the automated task impacts others, the interface should highlight the user's impact on those individuals.

In the workshop, participants discussed ways to exchange with peers about the oversight. One group included a 'Board of Shame' (\autoref{fig: bookOfShame}) in their prototype to share unique or humorous student answers. Others added student identifiers to facilitate discussions about specific solutions, even outside the interface.
Participants incorporated commenting features into automated grading to provide meaningful feedback and recognition. Further, they integrated sorting mechanisms and dedicated displays for critical cases to mitigate the risk of students failing due to AI errors.

\begin{enumerate}[labelindent=\parindent,leftmargin=*,label={\textbf{C8.}}]
    \item \textbf{Relationship to peers.} Consider how the interface can support the user in relating to other oversight persons.
\end{enumerate} 

\begin{enumerate}[labelindent=\parindent,leftmargin=*,label={\textbf{C9.}}]
    \item \textbf{Relationship to affected individuals.} Consider mechanisms that help users to recognize and positively shape the impact of their oversight on others.
\end{enumerate}
 
Interestingly, some participants expressed a desire to engage with the AI in relational terms, envisioning it as a collaborator with whom they could ``discuss'' recommendations, question decisions, and share responsibility, and gave it a human or animal-like appearance (\autoref{fig: relAsp}).  While SMART primarily conceptualizes relational work in human–human terms, these accounts suggest that similar psychological needs may extend to interactions with AI. Therefore, it is essential to consider how users interact with the AI they oversee. Users may satisfy their need for relatedness by engaging with AI in human-like ways, yet this risks bias, over-reliance, or emotional attachment~\cite{akbulut_all_2025, skjuve_longitudinal_2022, pentina_exploring_2023}. Designers must therefore consider how to shape AI appearance and interaction so users can meet relational needs without compromising oversight.
\begin{enumerate}[labelindent=\parindent,leftmargin=*,label={\textbf{C10.}}]
    \item \textbf{Relationship to AI.} Consider how the appearance and interaction with the overseen AI contributes to the user's relational needs but might affect oversight performance.
\end{enumerate}

\subsection{Tolerable work characteristics} 
\label{sec: tolerable}
Tolerable work assesses whether work requirements avoid overwhelming workload (low role overload), conflicting task requirements (low role conflict), or clashes with other responsibilities (low work–home conflict). Our participants' experiences indicated role overload and, possibly, role and work–home conflict.
Some participants experienced overload when manually grading all student submissions rather than overseeing the AI, a strategy recognized as inefficient but adopted, for example, from a sense of responsibility (\autoref{sec: appr+exp}). This approach may have caused role conflict, as participants’ self-imposed expectations (ensuring all grading was correct) clashed with their oversight role. The tension was intensified by the perception that responsibility for errors ultimately rested with them, to the extent that  they were willing to complete the task in their free time or on weekends, thereby creating tensions between work responsibilities and private life.
From a SMART perspective, these accounts illustrate how tolerable demands were not always achieved in the oversight task. Although the task was narrow in scope, attempts to balance efficiency, fairness, and responsibility sometimes made the demands feel difficult to manage. 

This balance underscores the need for designers to present user demands in a manageable way through the interface. Rather than overloading users, the interface should help them manage tasks efficiently and clearly distinguish between user and AI responsibilities.

In the workshop, participants added filtering and sorting functions and progress bars to help focus on achievable goals without becoming overwhelmed. Some groups also included reminders showing time spent and progress made to manage task load. To reduce role conflict and distinguish their oversight role from the AI’s grading, participants removed the ‘accept’ button for individual tests and, in some cases, refrained from making direct corrections.
\begin{enumerate}[labelindent=\parindent,leftmargin=*,label={\textbf{C11.}}]
    \item \textbf{Role Overload.} Consider how the interface can present the requirements of the oversight task in a manageable way.
\end{enumerate}
\begin{enumerate}[labelindent=\parindent,leftmargin=*,label={\textbf{C12.}}]
    \item \textbf{Role Conflicts.} Consider how the interface can clearly separate the tasks and responsibilities of the user from those of the AI.
\end{enumerate}

%% file: content/conclusion.tex
\section{Discussion and Conclusion}




This paper investigated effective and motivational human oversight of AI from a user-centered perspective. We analyzed experiences, discussions, and interface prototypes from four co-design workshops where domain experts were tasked to first oversee an automated AI grading system, then discussed their experiences and personal needs for oversight support, and finally reflected on these user requirements by creating concrete interface prototypes that could support them during AI oversight. Our thematic analysis revealed that domain experts often relied on approaches they later judged ineffective, leading to negative feelings towards both the AI and themselves. From participants' discussions, we inferred four user requirements for oversight support. Participants desired a clear understanding of their role, tasks, and responsibilities, as well as of the AI's capabilities and its decision process. They further wanted to contribute meaningfully to the task process, exchange with peers, and interact with the AI as a colleague. Through prototyping,  participants critically reflected on these requirements and discussed how to implement them in practice.

Importantly, we found that the identified user requirements aligned closely with the work design literature, which emphasizes the support of users’ psychological needs in interface design. Specifically, we integrated our findings with the SMART model of work design~\cite{parker2024smart} and proposed a comprehensive consideration framework for the design of user interfaces that support human oversight of AI. It centers around five general work characteristics, which we apply as characteristics of the interface: Stimulating, Mastery, Autonomous, Relational, and Tolerable interface characteristics. Each characteristic links the user requirements from our workshops with psychological processes found to impact long-term job satisfaction. Being rooted in a general theory of work design and informed by the task-specific outcome of our co-design workshops, we believe that this consideration framework will be applicable to a variety of domains where AI will increasingly be used to automate decision-making processes, such as in legal, medical, or educational domains, and many other. 

In the following, we discuss the trade-offs that should be considered when designing for different interface characteristics and how our proposed consideration framework relates to existing requirements for effective human oversight and to broader human-AI interaction guidelines. 

\subsection{Trade-offs and Task Constraints}
Our consideration framework  deliberatively avoids prescriptive guidelines, since designing for specific interface characteristics might entails trade-offs that may unintentionally affect other aspects of oversight or depend on the specific task and organizational context. When aiming to enhance particular work characteristics, designers should therefore remain mindful of potential tensions between considerations and carefully balance competing design goals.

For example, designing for a strong \textit{relationship with affected individuals} could increase an oversight person's \textit{role overload} or \textit{role conflict} if they find their responsibilities difficult to fulfill. 
Also, the quantity and nature of \textit{feedback} should be carefully considered to prevent \textit{role overload} or altering their \textit{role understanding}: if the overseer does not aim to re-evaluate all decisions, showing information about the number of tests they did not re-evaluate may overwhelm them or even alter their role understanding. 
Excessive \textit{autonomy} in task execution may similarly induce \textit{role overload} or hinder \textit{role understanding}.  Instead, designers may choose to restrict overseers in how they perform the oversight task, which they should, in turn, balance against users’ \textit{autonomy} and perception of the \textit{meaningfulness of oversight}. 


Interface designers should also consider that some characteristics of SMART were more relevant and observable through our workshops and can be considered for interface design more easily than others. Therefore, it is important to remember that all of these characteristics can be considered at multiple levels --- some are easier met at the interface level, while others need to be considered from the task or job level.
Sometimes the task itself may negatively affect a specific work characteristic, which can then be accounted for at the interface level, or the other way around. 
For example, 
for a mundane task where only very little information needs to be processed and interventions do not require critical thinking, challenge appraisal or work meaningfulness could be addressed through other \text{stimulating interface characteristics} or at the job level (e.g., combining oversight with other tasks). 
Interface designers should also consider that an oversight person’s \textit{relational} needs may be met outside their oversight task. When other work activities strengthen the \textit{relationship to peers}, relational features in the interface are less critical. However, when such opportunities are scarce, the interface should actively foster relatedness.
Different oversight tasks offer, for example, varying levels of \textit{feedback}. Some provide immediate responses to interventions, while others reveal outcomes only over time. In this case, it is important to consider that \textit{feedback} can also come from organizational sources, such as supervisors, or through training that enhances \textit{understanding of the oversight role and the AI’s capabilities}.
Lastly, work must be \textit{tolerable}. While interfaces can present workload in manageable ways, supervisors should avoid assigning unmanageable responsibilities. In high-load tasks, preventing \textit{role overload} by presenting information in a way that supports manageability becomes especially critical.

\subsection{Design Considerations for Effective Human Oversight}
While our consideration framework is mainly informed by the work design literature and findings from our co-design workshops, it also aligns with theoretical frameworks on effective human oversight.  Specifically, Sterz et al. argue that "an oversight person is effective in their human oversight if and only if they have causal power, epistemic access, self-control, and fitting intentions"~\cite[p.2499]{sterz_quest_2024}. 
According to this reasoning, effective oversight requires the ability to influence the system in ways relevant to one’s goals (causal power)~\cite{sterz_quest_2024}, a requirement reflected in our considerations: an oversight person who identifies with their role perceives their contribution as meaningful.
Furthermore, oversight personnel need sufficient knowledge of their decision situation, including the system, its state, potential interventions, and their consequences (epistemic access)~\cite{sterz_quest_2024}, which aligns with our considerations for the mastery work characteristics.
Sterz et al. further emphasize that effective oversight requires the oversight person to act autonomously and retain their attention (self-control)~\cite{sterz_quest_2024}. Similarly, our considerations highlight the importance of a stimulating task and ensuring the users' autonomy. 
Finally, the oversight person needs the intention to be effective (fitting intentions)~\cite{sterz_quest_2024}. While designers may not prevent people with conflicting interests or ill intentions from performing ineffective oversight, designing motivating and satisfying experiences could increase the likelihood of appropriate intentions.

Similar to our prior discussion, Sterz et al. argue that technical design (e.g., interface design), individual factors (e.g., training, domain expertise, motivation), and environmental factors (e.g., job design, accountability, time pressure) influence the extent to which these requirements are met. 
Overall, our considerations align with their requirements for effective human oversight but extend them by integrating work design principles and providing concrete guidance for interface designers.

\subsection{Extending Human–AI Interaction Guidelines for Human Oversight}
We also compare our design consideration framework with existing guidelines for human-AI interaction design (see~\cite{zhao_human-ai_2025} for an overview), since human oversight of AI could be considered a specialized form of human-AI interaction. 
Specifically, we reviewed three popular guidelines published by Google~\cite{google_google_2019}, Apple~\cite{apple_human_nodate}, and Microsoft Research~\cite{amershi_guidelines_2019}. We see an overlap in key aspects of our considerations, including promoting understanding of AI, granting autonomy, providing feedback, and preventing role overload and conflict. However, we also see that our considerations extend on these guidelines in three important ways.
First, we focused on the interaction between an AI that automates a task and a human overseeing its output. While participants encountered challenges, they were also optimistic and excited that the AI automated tasks they previously found meaningful and stimulating (Participant 3, Psy1: "In the end, I feel like it is still kind of exciting because maybe it does already kind of set a baseline"). Current guidelines recommend augmenting rather than automating stimulating tasks~\cite{google_google_2019}, yet in high-risk or critical work contexts, automation may be justified, for example, to enhance safety. Our considerations emphasize the importance of UI design to support users in understanding their role as an overseer and highlight other opportunities for experiencing work meaningfulness.
Second, existing guidelines address role conflict by recommending to adapt AI behavior to align with users’ initial behavior~\cite{apple_human_nodate, google_google_2019,amershi_guidelines_2019}. Our workshop revealed that participants initially experienced role conflict but resolved it by including specific UI elements and controls that would help them to adjust their own approach to the task (e.g., removing an Accept button). This underscores that role conflicts can be mitigated through changes in both AI and user behavior. 
Lastly,
current guidelines largely neglect users’ relational needs. Our findings highlight the value participants placed on relating to peers and those affected by their actions. These relational needs may be amplified by the specific task setting, suggesting that our considerations offer unique insights for designing oversight interfaces and that further research is needed to explore relational needs in other oversight contexts and human–AI interactions at the workplace.

\subsection{Limitations and Future Work}
A key limitation of our workshop concerns the characteristics of the oversight task. Domain experts graded a set of tests to evaluate the AI’s performance and detect potential errors, with no direct time pressure. 
While non-time-critical tasks are common in education, law, or healthcare, other domains, such as aviation, involve trained oversight experts or real-time interventions, which introduce different work characteristics. Consequently, the relevance of each consideration depends on the broader work context. For instance, in our grading task, participants had substantial autonomy in timing and method, making interface support for autonomy potentially less critical than in tasks with lower inherent autonomy. Future work should validate and potentially extend our considerations with requirements from other types of tasks.

Our participants had limited experience with automated grading tools and performed the task only briefly at the start of the workshop. In real work settings, oversight occurs over longer periods and repeated sessions, which may reveal additional problems or needs. Nevertheless, we expect our findings to generalize to longer interactions with AI-based decision-support systems, since prior research suggests that negative experiences and unmet psychological needs may intensify over time~\cite{faas_give_2024}.

Still, future work should investigate the user requirements of oversight personnel through longitudinal studies, across different domains, and also study individuals who regularly perform oversight. Another promising direction is to further examine the trade-offs between our considerations and provide more guidance on how to design good work for oversight personnel. Most importantly, though, more empirical work will be needed to evaluate the impact of specific considerations on users’ psychological processes and their satisfaction with the oversight task.

%% file: content/appendix.tex
\section{Example Task}
\label{app: task}
\textbf{Sample Solution}\\
Question 1:\\
Consider this declaration of bar. For which values of x and y does it terminate? For which does it diverge? Assume the ideal interpreter (i.e., an interpreter that can work with arbitrarily large integers without overflowing).\\
1 let rec bar (x: int) (y: int) : int = if x = 5 then y else bar (x - 2) y\\
Answer 1:\\
bar terminates iff x $\geq$ 5 and x is odd. For all even values of x and all x $<$ 5, bar diverges. The value of y does not impact the termination of bar\\
(1 point for termination condition, 1 point for divergence condition)\\
Achievable Points: 2\\
\textbf{Student Answer}\\
Student ID: 9\\
Answer 1:\\
It terminates whenever x $\geq$ 5 (because subtracting 2 repeatedly will eventually hit x = 5) and in that case returns y; it diverges whenever x $<$ 5. The value of y is irrelevant (any integer y is returned upon termination).\\
Achieved Points: 0.0

\section{Prompts}
\label{app:prompts}
We used the OpenAI o4-mini model to generate the 'student' answers for the 'Programming 1' exercises and to assign points to all student answers. To generate the students answers we prompted the OpenAI API with the following prompt: \\
"I have the following question for a test: \textit{Question}. There are \textit{Points} achievable points for this question. Generate an answer for this question, which a student would give that receives about \textit{Percentage}\% in a test. Please only provide the generated answer and no additional information." \\
For \textit{Question} we inserted the question for the given exercise and for \textit{Points} the amount of points one can achieve for this question. We prompted the OpenAI API 41 times for each question and varied the \textit{Percentage} so that it aligned with the distribution of achieved grades in the psychology exam (21 x 100\%, 15 x 75\%, 5 x 50\%).

For assigning points to the students' answers we used the following prompt structure:\\
"I have the following test question: \textit{Question} and the matching sample solution: \textit{Solution}. Achievable Points: \textit{Points}. This is my answer to the question: \textit{Answer}. How many points do I achieve with this answer to the question? Please only answer with the achieved points and do not provide any further information."\\
Here, we included the sample solution as \textit{Solution} and the students answer as \textit{Answer} additionally to the \textit{Question} and \textit{Points}. 

\section{Instructions}
\label{app:instructions}
\subsection{Session 1}
\paragraph{Purpose}
Due to current advancements in the field of artificial intelligence, there are more and more systems that support humans in their decision-making. Technologies have been developed that can compete with human decision-makers in all kinds of decision-making tasks. These systems are already in use for court decisions, hiring decisions, loan distribution, and more. Another domain where these technologies perform well is grading students' assignments, where they can support tutors and professors in their teaching.  While there are AI systems that perform well in these tasks, there are concerns about using the AI’s decisions without a human overseeing them. It is argued that humans are better at making decisions on edge cases or unusual situations, and also that humans employ ethical considerations in their decision-making. Based on such argumentation, many guidelines and laws require a human oversight person to oversee AI decision-making tools in high-stakes scenarios such as court decisions, loan distributions, and assessments of students’ performance.  In this workshop, we aim to identify the users’ needs, desires, problems, and strategies when overseeing an autonomous decision-making system and prototype interfaces for supporting these. Performing such a task can be very tiring, frustrating, and hard. AI systems are only used if they are very good at their task, which makes it very difficult for human oversight people to detect any errors made by the system. Evidence shows that the argument supporting human oversight is flawed and that people are not able to increase the accuracy of the AI systems. We want to investigate which goals oversight people can have, which of them are achievable and motivating, and how the user interface can support the user to have realistic goals and how to achieve them.
\paragraph{Task}
The workshop consists of two sessions with different goals. In the first session, you will be introduced to the task and have time to perform the task once. The task is to oversee an autonomous grading system that grades student tests. You will have 20 minutes to check if the autonomous system performed valid corrections or identify possible errors made by the system. After that, you will work in a group of people to reflect on and identify the specific strategies you used to approach the task and derive user needs and system requirements for an oversight interface. In the second session, you will brainstorm and select solutions for the needs and problems identified in the first session. After you have discussed different solutions, you will be provided with materials to create a paper mock-up of an interface that would support you in performing the grading oversight task. Both sessions will be moderated by an experimenter.
\paragraph{Collected Data}
We will video and audio record both sessions and take pictures of the mock-ups and diagrams created by you. 
The recorded videos are not anonymized. They will be stored on the Perspicuous Computing Nextcloud, which is a local server running at Saarland University and will only be accessible to researchers involved in this project. Videos will be deleted upon completion of the research project or if you request deletion of your data. We might use pictures from the workshop in future publications, but will ensure that participants’ faces are not recognizable. All other data will be pseudonymized.  No explicit clues about your identity will be left in the stored data. The anonymized data will be released as part of the publications on this research project.
\paragraph{Duration and Breaks}
Today’s session will take about 2 hours. You will have 20 minutes to perform the oversight task at the beginning. Then we will create an affinity diagram together in a group discussion. You can take breaks at any time. If you are feeling tired, ask for a break. It is very important that you can always give your best! Of course you are free to end the experiment at any time.
\paragraph{Task Instructions}
\label{app:task_instruction}
Imagine you are a Tutor of Programming 1 [Arbeits- und Organisationspsychologie]. Your job is to grade a test where 40 students participated, and which you will later hand out to the students. The test had four questions, and your Professor has given you the grading scheme. You are supported in this task by an automated grading system. It knows about the grading scheme and has assigned points to each student's answers. Due to the examination regulations of the university, a human oversight person is required to check if the autonomous system performed valid corrections or identify possible errors made by the system. Since the grading is already done by the automated system and you only need to check it, this should be faster than grading all the tests manually, and your Professor only allocated 20 minutes for all the tests. In the next 20 minutes, you have time to familiarize yourself with the task. Keep in mind that it is not the goal to perform the task as good as possible in these 20 minutes. Rather, you should use the time to think about how you would approach it, whether your initial strategy will be efficient, and if it is enjoyable to use this strategy. You should also think about other strategies for this task, and how efficient and enjoyable they would be. Think about different goals you could have when performing this task and how achievable they are, how much you would enjoy performing the task with this goal in mind, and which strategies would help you to achieve the goal.
\paragraph{Affinity Diagram}
\label{app:affinity_diagram}
Now we will create an affinity diagram to structure our thoughts and experiences. An affinity diagram is a visual tool used to categorize and structure large amounts of data or ideas. To create an affinity diagram, first, you will reflect on the strategies you took and the problems you faced when doing the task. You can write each thought on a sticky note. Try to use as few words as possible, so that everyone can read it from their seat. You will have enough time to explain your thoughts behind the keywords you wrote down. After collecting our thoughts individually, we will check each sticky note one by one and group them into categories, which are characterized by shared characteristics or relationships. Lastly, we will check for hierarchies and relations between the different categories.
\paragraph{Reflection Questions}
\label{app:reflection}
\begin{itemize}
    \item What were you trying to achieve while checking the AI grading?
    \item What are other goals you could try to achieve when checking the AI grading?
    \item Walk us through the steps you would take to achieve each goal
    \item How did you feel while checking the AI grading and what made you feel this way?
    \item If you were telling a friend about checking the AI grading, how would you describe it?
    \item What would you do differently the next time you have to check the AI grading?
    \item In what ways is checking the AI grading different than grading the tests yourself?
\end{itemize}

\subsection{Session 2}
\paragraph{Purpose}
In this session, we want to come up with ideas for a user interface that makes checking grades assigned by an AI easier and more fun. Remember the oversight task you performed in the last session. The goal of this session is to design a user interface prototype that would support you in performing this task in the future.
\paragraph{Task}
Today, in the second session, you will brainstorm and select solutions for the needs and problems identified in the first session. After you have discussed different solutions, you will be provided with materials to create a paper mock-up of an interface that would support you in performing the grading oversight task. 
\paragraph{Duration and Breaks}
The study will take about 120 minutes. You will have 30 minutes to come up with solutions and decide on the solutions you want to integrate into your prototype. Then you have 50 minutes to work on a paper mock-up of an oversight interface. You can take breaks at any time. If you are feeling tired, ask for a break. It is very important that you can always give your best! Of course you are free to end the experiment at any time.
\paragraph{Crazy 8's}
We will start this session with brainstorming solutions to make the oversight task from the previous session easier and more fun. We will use the crazy 8 method. This method is a brainstorming technique used to generate a large number of ideas in a short amount of time. It involves writing down 8 ideas within an 8-minute time frame, without stopping to think about their feasibility or practicality. The goal is to produce a large quantity of ideas, which can then be evaluated and refined later.